\documentclass[useAMS,usenatbib,usegraphicx]{mn2e}

\usepackage{amsmath}
\usepackage{graphicx}
\usepackage{longtable}

\title[Photodissociation in interstellar ices]{The efficiency of photodissociation for molecules in interstellar ices}
\author[J. Kalv\=ans et al.]{
J. Kalv\=ans\thanks{E-mail: juris.kalvans@venta.lv},\\
Engineering Research Institute "Ventspils International Radio Astronomy Centre" of Ventspils University College,\\
In$\check{z}$enieru 101, Ventspils, LV-3601, Latvia\\
}
\begin{document}

\date{Accepted 2018 April 24. Received 2018 April 11; in original form 2017 December 30}

\pagerange{\pageref{firstpage}--\pageref{lastpage}} \pubyear{201X}

\maketitle

\label{firstpage}

\begin{abstract}
Processing by interstellar photons affects the composition of the icy mantles on interstellar grains. The rate of photodissociation in solids differs from that of molecules in the gas phase. The aim of this work was to determine an average, general ratio between photodissociation coefficients for molecules in ice and gas. A 1D astrochemical model was utilized to simulate the chemical composition for a line of sight through a collapsing interstellar cloud core, whose interstellar extinction changes with time. At different extinctions, the calculated column densities of icy carbon oxides and ammonia (relative to water ice) were compared to observations. The latter were taken from literature data of background stars sampling ices in molecular clouds. The best-fit value for the solid/gas photodissociation coefficient ratio was found to be $\approx0.3$. In other words, gas-phase photodissociation rate coefficients have to be reduced by a factor of 0.3 before applying them to icy species. A crucial part of the model is a proper inclusion of cosmic-ray induced desorption. Observations sampling gas with total extinctions in excess of $\approx22$~mag were found to be uncorrelated to modelling results, possibly because of grains being covered with non-polar molecules.
\end{abstract}


\section{Introduction}
\label{1intro}

Interstellar ices are formed by accretion and processing of chemical species on dust grains in dense parts of molecular clouds. Processes on grain surface and in the ice allow the formation of many species, whose synthesis in the gas phase is inefficient, e.g., molecular hydrogen, water, carbon dioxide CO$_2$ and some complex organic molecules (COMs). The understanding of the chemical processing of interstellar ices is therefore essential in interpreting observations of dense clouds (and other objects) and clarifying the interstellar synthesis pathways of many chemical species.

The icy species are dissociated by the interstellar radiation field (ISRF) and cosmic-ray (CR) induced photons. The resulting molecular fragments may react with neighbouring molecules paving the way for a rich chemistry on the grain surface. These two steps -- dissociation and subsequent reactions -- constitute the phenomenon of photoprocessing of (interstellar) ices.

Ice photochemistry has a significant role in the synthesis of complex species in space \citep[e.g.,][]{Oberg11nasa}. The most direct method in determining the effects of UV photons on the composition of astrophysical ices is astrochemical laboratory experiments. Irradiation increases the number of different chemical species even in pure ices consisting of simple molecules \citep{Gerakines96,Bossa15,Martin15,Paardekooper16,Cruz18,Martin18}. Photolysis of icy mixtures considerably increases their chemical diversity \citep{Gerakines04,Oberg09aa2,Oberg10,Islam14,Jimenez14,Munoz14,Henderson15,Fedoseev16,Chuang17}.

The aim of this work is to evaluate the average efficiency of photodissociation for molecules in interstellar ices by comparing calculated and observed abundances of icy species. Such a study was permitted by the availability of observational data that relate the abundances of major icy species to total interstellar extinction $A_{\rm V}$ along a line of sight (LOS) towards a background star. By comparing the observed and calculated proportions of these species at certain $A_{\rm V}$(LOS) values, it is possible to deduce, to what extent ice composition has been affected by the photoprocessing of the icy mantle.

The main tasks corresponding to the above-mentioned aim are developing an astrochemical model suitable for investigating photochemistry in interstellar ices and obtaining results that can be reliably compared with observational data. The macrophysical model is described in Section~\ref{21phys}, while the chemical model is explained in Section~\ref{22chem}. Section~\ref{23phot} is dedicated on the approach for evaluating the photodissociation efficiency for icy species. Section~\ref{3res} describes the obtained results, with the final evaluation of the photodissociation efficiency discussed in Section~\ref{34epsilo}.

\section{Methods}
\label{2meth}

The chemical model `Alchemic-Venta' \citep{K15apj1} was employed. It is based on the tested ALCHEMIC engine \citep{Semenov10} and has been adapted for modelling contracting cloud cores. Attention was paid for improving model parts crucial to a ice photochemistry study with results compared to observations. These include the modelling of an entire LOS through the cloud core, newer data and more detailed description on photochemical processes in ice. The effects of CRs were revised in line with recent findings. In particular, cosmic-ray induced thermal desorption (CRD) plays a significant role in shaping the composition of the ices.

Because most of the observational data sample molecular cloud complexes, an embedded cloud core was considered. The observations (data and references listed in Appendix~\ref{app-obs}) sample a range of column densities, which were represented in the model in the course of the time-dependent core contraction.

In order to compare observed and calculated composition of interstellar ices, we employ the abundances of icy species along LOS, relative to water ice. Such an abundance ratio allows for a more reliable comparison with observational data than using absolute abundances (or abundances relative to H$_2$) because sight lines with similar $A_{\rm V}$(LOS) may sample different amounts of dense gas and ices. This means, for example, that the observed relative abundance $X_{\rm obs, CO}$ of solid carbon monoxide molecules is equal to the ratio of the observed CO ice and H$_2$O ice column densities $N_{\rm CO}/N_{\rm H_2O}$. If we employ column densities from the modelling results (at the respective $A_{\rm V}$(LOS)) in the latter relation, we obtain the calculated relative abundance of icy CO $X_{\rm calc, CO}$, which can be directly compared to $X_{\rm obs, CO}$.

\subsection{Physical model}
\label{21phys}
%
\begin{figure}
 \vspace{-1.0cm}
  \hspace{-1.0cm}
  \includegraphics[width=13.0cm]{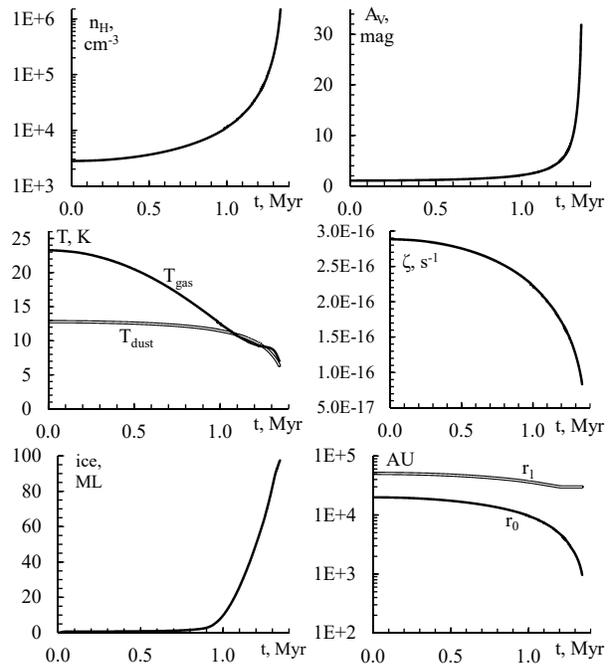}
	\vspace{-8.5cm}
 \caption{Evolution of key parameters at the centre of the modelled cloud core -- density $n_{\rm H,0}$, interstellar extinction $A_{\rm V}$, gas temperature $T_{\rm gas}$, dust grain temperature $T_{\rm dust}$, the CR ionization rate $\zeta$, indicative ice thickness (in monolayers) and cloud core parameters $r_0$ and $r_1$.}
 \label{att-phys}
\end{figure}
In order to obtain $X_{\rm calc}$, a chemical model of a contracting cloud core was employed. Relative abundances of  icy species were calculated for a LOS that goes through the centre of the core. As the central density increases with time, $A_{\rm V}$(LOS) rises, too. This allows to compare $X_{\rm obs}$ and $X_{\rm calc}$ at a range of $A_{\rm V}$(LOS) values.

The macrophysical model was created to reflect an approximately feasible density structure and previous chemical evolution of ices for each LOS corresponding to the observed $A_{\rm V}$(LOS) values in Table~\ref{tab-obs}. Of course, such an approach is not universal because the observed cores may have different parameters and history. In the course of this study, we identified several objects from the observational data that likely do not fit the description employed in the model (Sections \ref{31obs} and \ref{32lim}).

The above means that the chemistry was calculated for a one-dimensional model of the cloud core. This is a significant improvement over our previous 0D point models, and allows for a more precise comparison with observational data towards background stars. We opted for a core evolution scenario where the central density of the core $n_{\rm H,0}$ increases according to a delayed gravitational collapse. The cloud core starts with a $n_{\rm H,0}=2800$\,cm$^{-3}$, which grows according to Equation\,(1) of \citet{Nejad90} until a maximum density of $1.5\times10^6$\,cm$^{-3}$ is reached in an integration time $t=1.39$\,Myr. This corresponds to $A_{\rm V}$(LOS) from 2.1 to 63.7\,mag. Higher central densities were not considered because the maximum extinction along LOS reported in observational data is 56~mag.

The considered core has to be representative to some extent to most of the dense gas clumps encountered in the interstellar clouds. Many of the high values of $A_{\rm V}$(LOS) derived from the observations do not correspond to starless cores, indicating that these are heavily embedded or prestellar objects. Our chosen core can be described as a low-mass Bonnor-Ebert sphere \citep{Ebert55,Bonnor56,McCrea57}. The density profile for the sphere was expressed with a Plummer-like \citep{Plummer11,Whitworth01} approximation:
\begin{equation}
	\label{phys1}
	n_{{\rm H},r} = n_{\rm H,0} \left[1 + \left(\frac{r}{2.88r_0}\right)^2\right]^{-1.47},
\end{equation}
where $n_{{\rm H},r}$ is density at a given radius $r$, and $r_0$ is the radius of the central density plateau. The density profile for each time step for the 1D model was calculated with this equation. We did not consider events that trigger the contraction of the core.

The physical parameters of the core -- $n_{\rm H,0}$, $r_0$, total (outer) radius $r_1$ and core mass $M_{\rm core}=4.0$~$M_\odot$ -- were adopted from the rather diffuse yet supercritical core No.\,134 (hereinafter -- \#134) from Table~C.1 of \citet{Vaytet17}. The increase of $n_{\rm H,0}$ with time according to \citet{Nejad90} prompts changes in the core's density profile according to Equation\,(\ref{phys1}). The central density $n_{\rm H,0}=4.6\times10^4$~cm$^{-3}$ corresponding to \#134, is reached at integration time $t=1.21$~Myr.

The model follows the chemical evolution of gas parcels with a fixed mass coordinate along the LOS. In order to ensure that the absolute mass coordinate of each parcel is constant, the total mass of the core has to be constant. This was achieved by calculating the density plateau radius $r_0$ with a best-fit function in the form
\begin{equation}
	\label{phys2}
r_0 = C_1{n_{\rm H,0}}^{C_2}
\end{equation}
where $C_1$ and $C_2\approx-1/2$ \citep{Keto10,Taquet14} are parameters adjusted so that $M_{\rm core}$ always stays constant at 4.0~$M_\odot$. $r_0$ varies from 20~kAU at the start of the simulation to 970~AU at the end. We note that the model does not employ the free-fall times calculated by \citet{Vaytet17} because the core is already contracting when its parameters reach those of \#134. Slight changes of the core outer radius $r_1$ were also permitted in order to ensure constant core mass. The evolution of $r_0$ and $r_1$ are shown graphically in the lower-right plot of Fig.~\ref{att-phys}.

Our considered cloud core must not be isolated because the observations sample molecular cloud complexes. The core was assumed to be embedded in diffuse gas with isotropic column density $N_{\rm out}=5\times10^{20}$\,cm$^{-3}$. This value is slightly higher than that characteristic to isolated starless cores \citep[e.g.,][]{Lippok13} because the considered core is part of a larger complex. $N_{\rm out}$ cannot be much higher, otherwise the simulation would start with $A_{\rm V}$(LOS) close to 3\,mag, when an observable water ice layer should already be present on grain surfaces \citep{Whittet01}.

Furthermore, we assume that the core resides in a plane-parallel parent cloud that blocks the interstellar radiation from most directions reaching the sphere, i.e., the cloud core is irradiated only from two opposite sides along the LOS. This aspect is important when calculating the rates of ISRF-driven processes. Fig.~\ref{att-phys} shows the evolution of the physical conditions and indicative ice thickness in the centre of the cloud core. The central part contributes significantly to the ice mass observed along the LOS.

Theoretical models show that CR ionization rate $\zeta$ in molecular clouds depends on their column density \citep{Indriolo09,Padovani09,Chabot16}. Here we calculated $\zeta$ according to the 'High' model of \citet{Ivlev15}. Following \citet{Padovani13}, $\zeta$ was modified by a factor of 0.3 to account for magnetic reflection of CRs at column densities relevant to this model. Fig.~\ref{att-phys} shows that $\zeta$ values in the centre of the core change from $2.9\times10^{-16}$\,s$^{-1}$ to $8.3\times10^{-17}$\,s$^{-1}$. This approach is consistent with the CR energy spectrum used in calculating the whole-grain heating rate in Section~\ref{223desor}. The flux of CR-induced photons $F_{\rm CRph}$ was modified proportionally, assuming that $F_{\rm CRph}=4875$\,cm$^{-2}$s$^{-1}$ when $\zeta=1.3\times10^{-17}$\,s$^{-1}$ \citep{Tomasko68,Cecchi92}.

A standard flux of $1.0\times10^8$\,cm$^{-2}$\,s$^{-1}$ for ISRF photons was assumed. Self- and mutual shielding of H$_2$, CO and N$_2$ with the help of the tabulated data of \citet{Lee96} and \citet{Li13} were included. The calculation of the gas temperature of the cloud was adopted from \citet{K17}, while that of dust grains was calculated according to \citet{Hocuk17}.

The model considers olivine grains with radius $a=0.1$\,$\mu$m that constitute 1 per cent of cloud mass. The surface density of adsorption sites was taken to be $1.5\times10^{15}$\,cm$^{-2}$ \citep{Hasegawa92}. The assumed thickness of a single ice monolayer (ML) was $3.5\times10^{-8}$\,cm. Grain size and the number of surface adsorption sites $N_s$ are time-dependent, taking into account the number of accreted ice MLs and the corresponding ice thickness.

\subsection{Chemical model}
\label{22chem}
%
\begin{table}
 \centering
 \begin{minipage}{80mm}
\caption{Initial abundances of chemical species relative to H nuclei.}
\label{tab-ab}
  \begin{tabular}{@{}lc@{}}
  \hline
Species & Abundance \\
\hline
H$_2$ & 0.5 \\
He & 9.00E-2 \\
C & 1.40E-4 \\
N & 7.50E-5 \\
O & 3.20E-4 \\
F & 6.68E-9 \\
Na & 2.25E-9 \\
Mg & 1.09E-8 \\
Si & 9.74E-9 \\
P & 2.16E-10 \\
S & 9.14E-8 \\
Cl & 1.00E-9 \\
Fe & 2.74E-9 \\
\hline
\end{tabular}\\
\end{minipage}
\end{table}
\subsubsection{Chemical reactions network}
\label{221netw}
The modelled cloud core starts with hydrogen in molecular form and other elements in neutral atomic form with initial abundances listed in Table~\ref{tab-ab}. For modelling gas-phase chemical processes, the UMIST Database for Astrochemistry 2012 \citep[UDfA12,][]{McElroy13} was employed. Photoreactions for icy species were also adopted from UDfA12. However, this database does not contain binary surface reactions. These were adopted from the COMs chemistry network of \citet{Garrod08} with changes from \citet{Laas11} and \citet{K15apj2,K15apj1}. The energy barrier $E_{\rm A,O+CO}$ for the surface reaction $\rm O+CO\rightarrow CO_2$, which is important for the production of CO$_2$, was taken to be 630~K according to the experimental results of \citet{Minissale13}. 630~K also happens to be a best-fit value for $E_{\rm A,O+CO}$, which allows to obtain results that are more consistent with observations than values used previously -- 290~K \citep{Roser01} or 1000~K \citep[see also \citeauthor{Garrod11} \citeyear{Garrod11}]{DHendecourt85}.


Surface binary reactions were described with the modified rate equations method, as in the original ALCHEMIC code \citep{Caselli02,Garrod09}. We included the diffusion-reaction competition following \citet{Garrod11}. Reaction and diffusion barriers can be overcome either by thermal hopping or quantum tunnelling, depending on which is faster.

The desorption energies $E_{\rm D}$ for surface species were adopted from the COMs network \citep{Garrod06,Garrod08}. These authors derived $E_{\rm D}$ for molecules on water ice surface. Their $E_{\rm D}$ values roughly agree with experimental values for the species relevant in this study \citep[e.g.][]{Martin14,Fayolle16}. However, for CO-dominated ices that appear late in the evolution of the cloud core, desorption energy and the associated binding energy (see below) can be notably lower, as indicated by thermal desorption of pure ices \citep{Collings04,Luna17}. This aspect was taken into account when comparing modelling results with observational data (Section~\ref{32lim}).

\subsubsection{Model of the icy grain mantle}
\label{222mant}
%
\begin{table*}
 \centering
 \begin{minipage}{160mm}
\caption{Summary of the energies characterizing the mobility icy species on the surface and in the bulk ice.}
\label{tab-en}
  \begin{tabular}{@{}llcllc@{}}
  \hline	
\multicolumn{3}{c}{Surface} $|$& \multicolumn{3}{c}{Bulk ice (mantle sublayers)} \\
Energy & Notation & Value & Energy & Notation & Value \\
  \hline
Desorption (adsorption) & $E_{\rm D}$ & (base data)$^{\rm a}$ & Absorption & $E_{\rm B}$ & 3.0$E_{\rm D}$$^{\rm c,d}$ \\
Binding (diffusion) & $E_{\rm b,s}$ & $0.5E_D$$^{\rm b}$ & Binding (diffusion) & $E_{\rm b,m}$ & 0.5$E_{\rm B}$ \\
. . . & . . . & . . . & Proximity & $E_{\rm prox}$ & 0.1$E_{\rm B}$$^{\rm d}$ \\
  \hline
\end{tabular}\\
$^{\rm a}$\,Surface chemistry network of \citet{Garrod08}; see text.\\
$^{\rm b}$\,\citet{Garrod06}. \\
$^{\rm c}$\,\citet{K15aa}. \\
$^{\rm d}$\,\citet{K15apj1}.
\end{minipage}
\end{table*}
The description of the icy grain mantle was adopted from \citet{K15apj1}. The mantle is formed on the grain surface via the adsorption of neutral species. The mantle was described as consisting of four layers -- the surface and three bulk-ice layers. The sticking coefficient for H and H$_2$ was adopted from \citet{Thi10}, while that of other species was taken to be unity. Species in the bulk ice were assumed to be bound with absorption energy $E_{\rm B}$ that is three times higher than the adsorption energy $E_{\rm D}$ for surface species.

Diffusion occurs between adjacent ice mantle layers. The diffusion energy for species within the mantle $E_{\rm b,m}$ was taken to be three times higher than that of surface species $E_{\rm b,s}$ \citet{K15aa}. The rates of chemical reactions for molecules in the bulk-ice layers were calculated by assuming that the molecule vibrates in its ice lattice cell. Because the cell is small, the molecule in consideration approaches its neighbours (potential reaction partners) with its characteristic vibration frequency $\nu_0$. A reaction occurs once the vibrating molecule happens to overcome a proximity barrier $E_{\rm prox}$ \citet{K15apj1}. This is opposed to the diffusion process over a significant surface area, which is the case of surface species \citep{Hasegawa92}. 

The relations between energies related to icy molecular processes on grains are summarized in Table~\ref{tab-en}. We refer the reader to \citet{K15aa,K15apj1} for more details, discussion and justification on the chemistry of the icy mantle. 

A conservative value of 0.50 for the binding to desorption energy ratio was adopted, as in the original COMs surface chemistry network of \citet{Garrod06}. Values such as 0.3, 0.35, 0.40, 0.50, 0.55 and 0.77 have been used for the $E_{\rm b}/E_{\rm D}$ ratio \citep[][respectively]{Hasegawa92,Garrod13,Garrod11,Garrod06,Vasyunin17,Ruffle01b}. Test calculations were performed with all these $E_{\rm b}/E_{\rm D}$ ratios, with the middle value of 0.50 yielding results that are most consistent with the `valid' observations, summarized in Section~\ref{32lim}.

\subsubsection{Desorption mechanisms}
\label{223desor}
%
\begin{table}
 \centering
 \begin{minipage}{80mm}
\caption{Adopted total desorption yields for surface chemical species in the model.}
\label{tab-yield}
  \begin{tabular}{@{}lccl@{}}
  \hline
Molecule & $Y_{\rm ISRF}$ & $Y_{\rm CRph}$ & References \\
\hline
N$_2$ & 5.5E-03 & 3.0E-03 & (1) \\
O$_2$ & 9.3E-04 & 9.3E-04 & (2) \\
CO & 5.7E-03 & 3.9E-03 & (1) \\
CH$_4$ & 2.0E-03 & 2.2E-03 & (3) \\
CO$_2$ & $f(B)$* & $f(B)$* & (4) \\
CH$_3$OH & 2.3E-04 & 2.5E-04 & (5) \\
H$_2$O & $f(B,T_{\rm dust})$* & $f(B,T_{\rm dust})$* & (6) \\
NH$_3$ & 2.0E-03 & 0.0E-03 & (7) \\
Other & 1.0E-03 & 1.0E-03 & . . . \\
\hline
\end{tabular}\\
*\,Function of ice thickness $B$ (ML) and grain temperature $T_{\rm dust}$. \\
References: (1) \citet{Bertin13}; (2) \citet{Martin15}; (3) \citet{Dupuy17}; (4) \citet{Oberg09aa1}; (5) \citet{Bertin16}; (6) \citet{Oberg09apj}; (7) \citet{Martin18}.
\end{minipage}
\end{table}
The model considers several desorption mechanisms that transfer molecules from the grain surface layer to the gas phase. These include evaporation, photodesorption by the ISRF and CR-induced photons, reactive desorption \citep{Garrod07}, and CRD. We also included the encounter desorption from \citet{Hincelin15}. The latter ensures that H$_2$ does not accumulate on grains at temperatures lower than $\approx9$\,K.

\citet{K15apj1} found that a selective desorption mechanism helps removing the over-abundance of icy carbon oxides CO and CO$_2$ in time-dependent astrochemical simulations. The efficiency of such a mechanism should be inversely dependent on the $E_{\rm D}$ of the surface species. It was suggested that indirect reactive desorption by H atoms combining on grain surface may provide such a mechanism. However, little data is available about this mechanism beyond theoretical estimates and it was not included in the present model. Instead, recent data indicates that this role can be played by CRD. The ability of energetic ions to induce desorption of icy species has been shown experimentally \citep[e.g.,][]{Johnson91,Dartois15} and studied theoretically \citep[e.g.,][]{Bringa04,Herbst06}. Such ions may induce also other processes in ices \citep[and references therein]{K15aa}.

\citet{Chabot16} and \citet{K16} showed that CRD-inducing whole-grain heating events happen more often than previously thought \citep[see][]{Hasegawa93,Roberts07}. Moreover, CR intensity (and thus, frequency of grain heating) depends on the column density of gas \citep{Padovani09}.

In this paper, the basic approach on CRD was retained from Equation\,(15) of \citet{Hasegawa93}, where it was assumed that CRD occurs for grains that are heated to 70~K for a cooling time $t_{\rm cool}=10^{-5}$~s. However, the frequency $f_{70}$ of grain encounters with CRs heating the grains to 70~K was recalculated using the data and methods of \citet{K16} as explained in Appendix~\ref{app-crd}. Therefore, the CR-induced desorption rate coefficient is
   \begin{equation}
   \label{chem1}
k_{\rm crd}(i) = k_{\rm evap,}(i, 70\,\rmn{K}) t_{\rm cool}(70\,\rmn{K})\times f_{70},~s^{-1},
   \end{equation}
where $k_{\rm evap,}(i, 70\,\rmn{K})$,\,s$^{-1}$ is the evaporation rate coefficient for species $i$ at 70\,K and $t_{\rm cool}(70\,\rmn{K})\times f_{70}$ is the 'duty cycle' or fraction of time spent in the vicinity of 70\,K for the grain. The frequency of CRD events $f_{70}$ is defined with Equations (\ref{crd1}) and (\ref{crd2}) in Appendix~\ref{app-crd}.

\subsubsection{Photoprocessing of icy species}
\label{224surf}
The model considers ISRF and CR-induced photodissociation of species in all ice layers. The rate coefficients were taken from the corresponding gas-phase reactions and modified according to
\begin{equation}
	\label{chem2}
	k'_{\rm ph} = \frac{1}{2}\epsilon_{\rm ph}k_{\rm ph}(1-P_{\rm abs})^B,~s^{-1}
\end{equation}
where $k'_{\rm ph}$ is the actual rate coefficient used in the model, $k_{\rm ph}$ is the dissociation rate coefficient for gaseous species \citep[Equations (3) and (4) in][]{McElroy13}, $P_{\rm abs}$ is photon absorption probability per ice ML  and $B$ number of ice MLs between the outer ice mantle surface and the molecule in consideration. The efficiency of photodissociation for icy species $\epsilon_{\rm ph}$ is defined with Equation\,(\ref{phot1}) below. The factor 1/2 was added to reflect the fact that a molecule residing on grain surface is partially shielded from incident radiation by the body of the grain, much larger than the molecule.

The value for $P_{\rm abs}=0.007$ was derived from molecular dynamics simulations \citep{Andersson08}. This means that, for a 100\,ML thick mantle, molecules in the bottom layer (adjacent to grain nucleus) would be dissociated with a rate reduced by a factor of $\approx0.5$. The $P_{\rm abs}$ value of 0.007 can be compared to values derived from experimental dissociation cross-sections of water, where $P_{\rm abs}\approx0.002$ \citep{Mason06,Cruz14i}. The higher, theoretically obtained value of $P_{\rm abs}$ was employed here to account for photon losses in processes other than full molecular dissociation, which is the process measured in the experiments.

Experiments show that photodissociation and desorption (photodissociative desorption, PDD) of molecules in the surface layer occur simultaneously \citep{Fillion14,Bertin16,Cruz16,Martin16}. In the case of amorphous water ice surface photodissociation, the proportion of water molecules going to the gas phase is about three per cent. A significant fraction of these desorbed molecules are dissociated \citep{Andersson08,Arasa15}. In astrochemical modelling, PDD has been accounted by \citet{Taquet13} for the water molecule. In order to consider PDD at least to some extent for all surface molecules, we employed a general approach on PDD, which is described in Appendix~\ref{app-pdd}.

The total rate coefficient for photodesorption (intact molecules plus their dissociated fragments) is calculated with
   \begin{equation}
   \label{chem3}
k_{\rm{pd}}= \frac{1}{2}\times\frac{\sigma_{\rm g} F_{\rm ph} Y_{\rm ph}}{N_s},~\rm s^{-1}
   \end{equation}
where $\sigma_{\rm g}$ (cm$^2$) is the grain cross section, $N_s$ is the number of adsorption sites on grain surface, $F_{\rm{ph}}$ is the flux (s$^{-1}$cm$^{-2}$) of either ISRF or CR-induced photons, and $Y_{\rm ph}$ is the total desorption yield either for interstellar ($Y_{\rm ISRF}$) or CR-induced photons ($Y_{\rm CRph}$).

Experimentally determined total desorption yields were adopted for CO, CH$_4$, CO$_2$, CH$_3$OH, N$_2$, NH$_3$, O$_2$ and H$_2$O. For other surface species, $Y_{\rm ph}$ was taken to be $1.0\times10^{-3}$. The adopted values of the total yields, summarized in Table~\ref{tab-yield}, include both, intact molecule desorption and desorption of photodissociation fragments. There is a variety of available experimental photodesorption yields for some common icy species, such as carbon monoxide \citep{Oberg07,Oberg09aa1,Munoz10,Munoz16,Fayolle11,Bertin13,Chen14,Paardekooper16}, carbon dioxide \citep{Oberg09aa1,Fillion14,Martin15}, methanol \citep{Oberg09aa2,Bertin16,Cruz16,Martin16} and nitrogen \citep{Oberg09aa1,Bertin13,Fayolle13}. Whenever possible, we employed $Y_{\rm ph}$ values from experiments with mixed ices, which presumably more closely represent the actual icy mantles in the ISM.

\subsection{Investigating ice photodissociation efficiency}
\label{23phot}

The model described above was used to obtain results to be compared with abundances of icy species along observed LOS ($X_{\rm calc}$ and $X_{\rm obs}$). We remind that at the core of this study is estimating the molecular photodissociation rate in interstellar ices. Photoprocessing of surface species was introduced in astrochemical modelling by \citet{Ruffle01a}. They assumed that dissociation of icy species occurs with an efficiency equal to that of gas-phase species. This is also the approach currently used in most astrochemical models considering surface chemistry.

Some of current astrochemical models that consider layered ices do not apply photoprocessing for bulk-ice molecules, often because of computational reasons \citep[e.g.][]{Taquet14,Ruaud16,Sipila16,Furuya17}. Others include full photoprocessing of the whole ice mantle \citep{Chang16,K17,Garrod17,Vasyunin17}. In this paper, we assume that photodissociation occurs with similar efficiency for surface and bulk molecules. However, for this core collapse model, the photodissociation of molecules in the \textit{surface} layer is what affects most the calculated column densities of icy species. The photoprocess in the model is dominated by the ISRF photons, whose flux is much higher than that of CR-induced photons for most of the integration (and ice formation) time.

Following \citet{Mason06} and \citet{Furuya13}, the ice molecule photodissociation efficiency $\epsilon_{\rm ph}$ can be expressed with
\begin{equation}
	\label{phot1}
	\epsilon_{\rm ph} = \sigma_{\rmn{ph, ice,} i}/\sigma_{\rmn{ph, gas,} i} ,
\end{equation}
where $\sigma_{\rmn{ph, ice,} i}$ is the photodissociation cross section for a given species $i$ in solid phase and $\sigma_{\rmn{ph, gas,} i}$ is the cross section for the same species in the gas phase, usually a known quantity.

An assortment of experimental data on the photodissociation cross sections of species in interstellar ice analogues are available. We reviewed the data provided by \citet{Gerakines96,Cottin03,Mason06,Oberg09aa2,Oberg10} and \citet{Cruz14i,Cruz14ii,Cruz16}. Some of these data have been summarized in Table~1 of \citet{Oberg16}. These experimental photodissociation cross sections were compared to ISRF photon cross sections of the respective gas-phase species derived from the UDfA12 database. It was found that, for most cases, the efficiency $\epsilon_{\rm ph}$ lies in the range of 0.03...0.72.

The above-mentioned experiments employ a variety of set-up approaches -- temperature of ice freezing and during irradiation, pure ices or mixtures, substrate materials, irradiation UV spectra, quantification techniques of chemical species. In some cases, different experiments considering the same species (e.g., CO, CH$_4$) yield cross sections that differ by up to two orders of magnitude. Because of this, the most reasonable approach in this study was to adopt a single $\epsilon_{\rm ph}$ value for all photodissociation reactions. This approach also has the benefit of being simple and easily applicable in astrochemical models.

In order to obtain an estimate of $\epsilon_{\rm ph}$, we performed the 1D simulations with 14 values of $\epsilon_{\rm ph}$ that cover the range indicated by experiments (Table~\ref{tab-epsilo}). For benchmarking, simulations with $\epsilon_{\rm ph}=0$ and $\epsilon_{\rm ph}=1$ were also included (a total of 1.6 million integration steps).

\section{Results}
\label{3res}

\subsection{Selection of observational data}
\label{31obs}

Several simple rules had to be obeyed when selecting the observational data. First, the observations have to sample quiescent medium, which means that the focus is on background stars. Second, the $A_{\rm V}$(LOS) value must be derived; this means that a number of earlier observational studies were not useful for our purposes. Third, the column density (or abundance) of water ice must be known, so that it is possible to obtain $X_{\rm obs}$. The fourth and final aspect was that the column density of a second species, besides water, must be reported in the paper.

In total, we selected 77 sources (from 9 published papers) that conformed to the above rules. These are listed in Appendix~\ref{app-obs}. Multiple observations sampling the same object were permitted (e.g., Elias~13, Elias~16). 18 of the sources are identified by authors \citep{Shuping00,Boogert13} being in proximity of high-mass star-forming regions, thus likely receiving irradiation doses significantly above to the average interstellar medium. This means that the model presented in Section~\ref{2meth} may be far from reflecting the conditions in these clouds. We included these irradiated samples in Appendix~\ref{app-obs}, figures and the ensuing discussion but they were not considered in deriving the final result of this investigation.

\subsection{Limitations of the calculated data}
\label{32lim}
%
\begin{figure*}
  \includegraphics{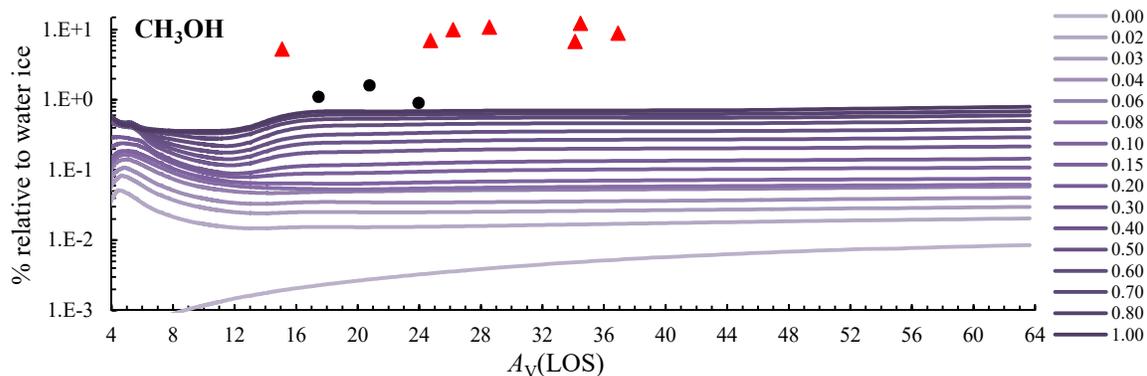}
	\vspace{-23.5cm}
 \caption{Comparison of calculated (curves with indicated $\epsilon_{\rm ph}$) and observed (circles) CH$_3$OH:H$_2$O ice column density ratios as a function of $A_{rm V}$(LOS). Red triangles (CH$_3$OH:H$_2$O exceeding 5 per cent) are suggested to represent observations of long-lived stable starless cloud cores.}
 \label{att-ch3oh}
\end{figure*}
%
In the numerical study, a single cloud core model was employed, while the observations sample multiple cores with basically unknown parameters. Because of this, it can be expected that not all observations can be used in this study. When comparing the observational data with the results of the calculations, it was realized that there exist two populations of observational data points that cannot be reproduced with the model.

The first set of such data points is sight lines with abundant interstellar solid methanol. This model is able to produce icy methanol with $X_{\rm calc, CH_3OH}$ of 1.5 per cent at maximum (in simulation with $\epsilon_{\rm ph}=1.00$), while \citet{Whittet11} report $X_{\rm obs, CH_3OH}$ of up to 12 per cent (Fig.~\ref{att-ch3oh}). Such abundances can be reproduced for interstellar cloud cores that are stable and exist for hundreds of kyr \citep{K15apj2}. Our modelled contracting prestellar core does not fit this description. Therefore, we assume that observations with $X_{\rm obs, CH_3OH}$ notably exceeding one per cent (5 per cent or more in this case) sample such long-lived cores and cannot be used for comparison with calculations. The respective data points are marked as red triangles in the figures.

The second data set, unsuitable for comparison with calculations, are observations sampling ices with $A_{\rm V}$(LOS) values exceeding $\approx22$~mag. We were unable to achieve a good agreement for data points ($X_{\rm obs}$ for CO, CO$_2$ and NH$_3$) with $A_{\rm V}$(LOS) exceeding this value within reasonable values for the parameters $E_{\rm b,s}/E_{\rm D}$, $E_{\rm A,O+CO}$ and $\epsilon_{\rm ph}$. The reason for such a systemic deviation could be either that the parameters of the modelled cloud core deviate significantly from the observed objects, or, more likely, that the properties of the grain surface changes when the surface gets completely covered with CO and N$_2$ as suggested by \citet{Bergin95}. Such a coverage with non-polar molecules reduces the adsorption and binding (diffusion) energies of surface species, affecting their chemistry. Easier diffusion of CO and other heavy reactants (e.g., O atoms) may allow them to be more competitive to H atoms, increasing the formation rate of CO$_2$ and reducing that of NH$_3$. This suggestion is based on simulations with lowered $E_{\rm b,s}/E_{\rm D}<0.5$ ratios, mentioned in Section~\ref{222mant}. A proper quantification and inclusion in the model of such a surface transformation is worth a separate study and was not attempted here. \citet{Cazaux17} offer a recent experimental-theoretical study on this topic.

Summarizing the above, the following observational data points are excluded from estimating the value of $\epsilon_{\rm ph}$ in Section~\ref{34epsilo}:
\begin{itemize}
	\item observations sampling regions with increased irradiation from nearby massive stars, as indicated by the authors of the observational studies;
	\item observations associated with long-lived starless cores, judging by their high abundance of methanol ice;
	\item observations sampling regions with $A_{\rm V}$(LOS) exceeding 22~mag.
\end{itemize}
Fig.~\ref{att-ices} shows the modelled curves of $X_{\rm calc}$ and $X_{\rm obs}$ points from literature data for CO, CO$_2$, NH$_3$ and N$_2$. The observational data excluded from the evaluation of $\epsilon_{\rm ph}$ are indicated in the figure.

\subsection{Photochemistry of major interstellar icy species}
\label{33phchem}
%
\begin{figure*}
  \hspace{-1.0cm}
  \includegraphics{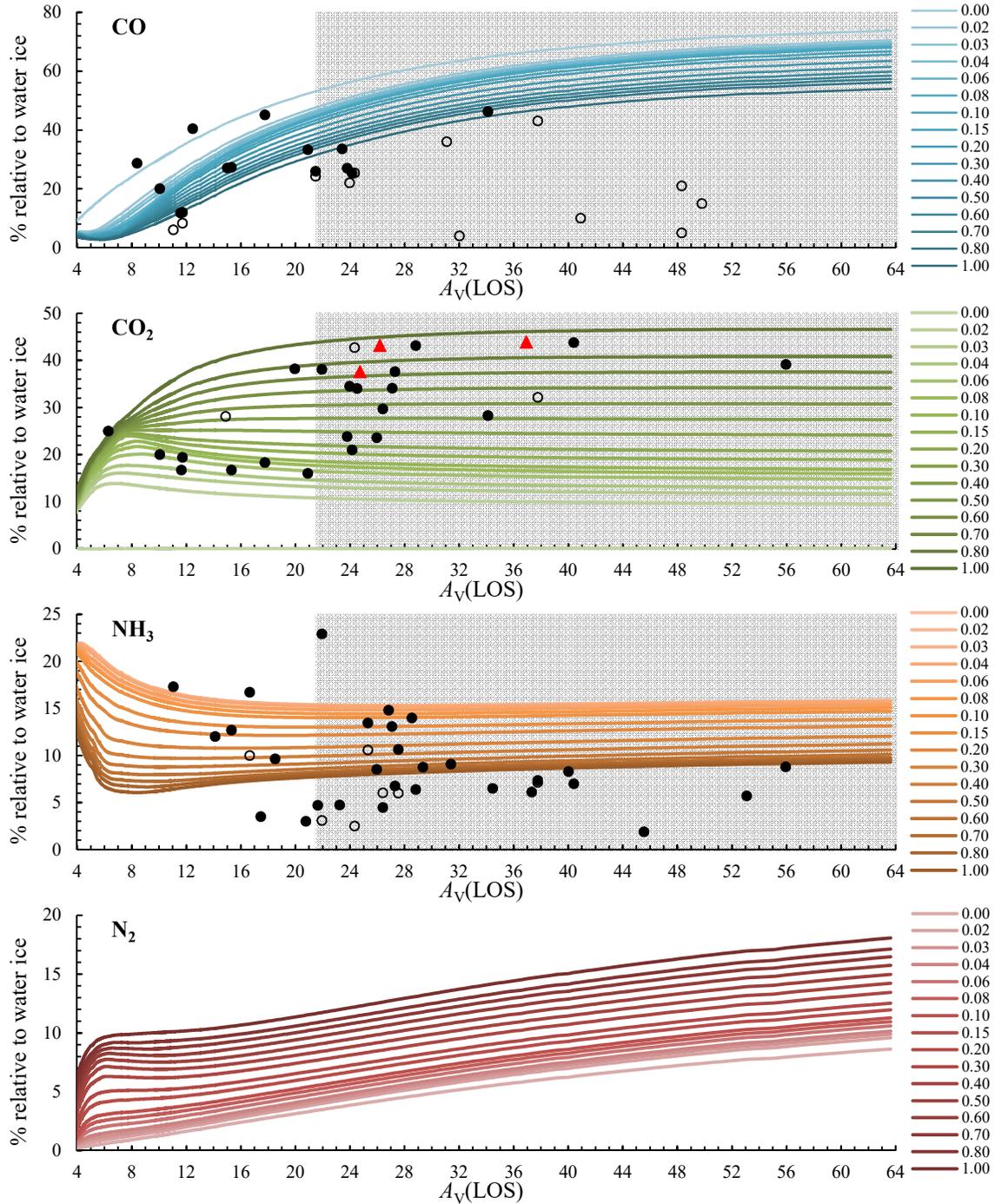}
	\vspace{-10.0cm}
 \caption{Modelled $X_{\rm calc}$ for CO, CO$_2$, NH$_3$, and N$_2$ ices relative to water ice (curves; from top to bottom) for simulations with indicated values of $\epsilon_{\rm ph}$ as functions of total interstellar extinction along the line of sight. These are compared to the observed abundances $X_{\rm obs}$, described as follows. Filled circles: `valid' data points in the context of this study; empty circles: observations sampling highly irradiated molecular clouds, as indicated in Table~\ref{tab-obs}; red triangles: observations attributed to long-lived starless cores. The shaded area indicates $A_{\rm V}\rm (LOS)>22$~mag, where observational data was found to significantly deviate from calculations (see Section~\ref{32lim}). In the case of CO$2$, several of the empty circles are outside the scope of the ordinate.}
 \label{att-ices}
\end{figure*}
%
The most abundant observed interstellar icy species are water and carbon oxides. The most important photo-chemical process in ices in the interstellar medium (ISM) is the photo-production of icy CO$_2$ at the expense of CO and H$_2$O \citep{Whittet98}. This study also investigated the effect of photoprocessing on the abundances of interstellar icy ammonia NH$_3$ and methanol CH$_3$OH \citep[e.g.,][]{Boogert15}.

While the choice of $\epsilon_{\rm ph}$ affects all icy species, there are only a few photoreactions, known and included in the network, that truly contribute to the abundances of the observed ice molecules. Before deriving the value of $\epsilon_{\rm ph}$, we briefly discuss the photochemistry of the main icy species.

The most abundant interstellar ice molecule is water, whose calculated and observed column density was used as a benchmark to obtain the relative abundances of CO, CO$_2$, and NH$_3$. The main formation route of water is the hydrogenation of O atoms on grain surfaces. The dissociation of water is the most important ice photoreaction, for which $\epsilon_{\rm ph}$ we estimated to be 0.43 and 0.72 from the data of \citet{Mason06} and \citet{Cruz14i}, respectively. Photoprocessing tends to reduce the abundance of water ice by converting it to O$_2$, H$_2$O$_2$ or, in the presence of CO, to CO$_2$ and CH$_3$OH. At least some the residual hydrogen escapes to the gas phase \citep{K10,Cruz18}. In addition to reduced dissociation rates, the UV absorption spectra is shifted for solid species when compared to gas. This was not explicitly considered in the study.

As a gas-phase product, carbon monoxide ice arises via accretion on grain surfaces. CO ice can be either hydrogenated to CH$_3$OH or oxidized to CO$_2$. For the conditions considered in this study, only the latter process is of significant importance. In ISM regions with a higher flux of UV radiation (empty circles in Fig.~\ref{att-ices}), the observed abundance of CO ice relative to water ice is notably lower than that obtained by the calculations. This agrees well with the picture that irradiation converts CO:H$_2$O icy mixtures into CO$_2$.

Photochemistry is crucial for the formation of CO$_2$ -- when $\epsilon_{\rm ph}=0$, the calculated column density of CO$_2$ relative to that of water ice is negligible, below 0.1 per cent. This means that all the CO$_2$ in our model is produced via photoprocessing of surface species. O and OH radicals are split from CO and H$_2$O before combining with CO molecules to produce CO$_2$. Observational data shows that cloud cores under increased irradiation generally have a higher CO$_2$:H$_2$O ice abundance ratio, even exceeding 100 per cent. A similar effect can be observed for the long-lived cloud cores (identified here by the high abundance of methanol ice), where the ice has been exposed to irradiation probably for hundreds of kyr \citep{K15apj2}.

The direct correlation between CO$_2$ and photoprocessing also means that the inclusion of a full LOS with the 1D model increases the calculated CO$_2$:H$_2$O ratio. This is because ices in the outer part of the cloud core are more exposed to the ISRF. For example, at the end of simulation ($A_{\rm V}\rm{(LOS)}=64$~mag) with $\epsilon_{\rm ph}$ taken to be 0.30, CO$_2$:H$_2$O at the very centre of the core is 19 per cent, while for the whole LOS it is 24 per cent.

While NH$_3$ ice is formed via surface reactions and N$_2$ is simply accreted from the gas, their surface chemistry is intertwined. Because N$_2$ has smaller cross-section for ISRF-induced photodissociation ($1.35\times10^{-18}$~cm$^2$ versus that of $7.05\times10^{-18}$~cm$^2$ for NH$_3$) and is N$_2$ shielded in the gas phase, any overall increase of photodissociation rates favours N$_2$. This is also manifested in Fig.~\ref{att-ices}, where the abundance of N$_2$ ice grows and that of NH$_3$ ice is reduced by increasing values of $\epsilon_{\rm ph}$. This is primarily because the photodestruction of NH$_3$ ice on the surface means that the remaining radicals must undergo binary reactions to recombine (e.g., $\rm NH + H$, $\rm NH_2 + H$) and in each such step some molecules are lost to the gas phase (where they are destroyed) because of reactive desorption. The observational data seemingly confirm these theoretical findings: cloud cores under increased irradiation tend to show lower relative abundances $X_{\rm obs}$ of NH$_3$ ice (Fig.~\ref{att-ices}).

\subsection{Calculation of ice photodissociation efficiency}
\label{34epsilo}
%
\begin{table}
 \centering
 \begin{minipage}{80mm}
\caption{Total difference between observed and calculated LOS abundances for CO, CO$_2$ and NH$_3$ ices relative to water ice.}
\label{tab-epsilo}
  \begin{tabular}{@{}lccc@{}}
  \hline
 & \multicolumn{3}{c}{$\sum (X_{\rm obs}-X_{\rm calc})$, percentage points} \\
$\epsilon_{\rm ph}$ & CO & CO$_2$ & NH$_3$ \\
\hline
0.00 & -183.1 & 208.2 & -50.6 \\
0.02 & -95.9 & 101.9 & -56.7 \\
0.03 & -79.2 & 80.0 & -56.5 \\
0.04 & -68.8 & 65.8 & -55.0 \\
0.06 & -56.9 & 48.2 & -50.7 \\
0.08 & -49.4 & 36.9 & -45.8 \\
0.10 & -44.8 & 29.6 & -40.6 \\
0.15 & -35.5 & 16.5 & -28.6 \\
0.20 & -26.0 & 5.4 & -18.5 \\
0.30 & -8.4 & -13.7 & -3.3 \\
0.40 & 7.9 & -31.5 & 7.9 \\
0.50 & 23.4 & -49.2 & 17.0 \\
0.60 & 37.1 & -67.7 & 24.1 \\
0.70 & 48.6 & -86.4 & 29.0 \\
0.80 & 59.6 & -105.4 & 32.4 \\
1.00 & 80.8 & -136.4 & 35.7 \\
\hline
\end{tabular}\\
\end{minipage}
\end{table}
Before finally deriving the value of $\epsilon_{\rm ph}$, we note that astrochemical models consider thousands of molecular processes with approximate rates, and a number of free parameters. The impact of uncertainties in these values can be significant and has been assessed in a number of studies \citep[e.g.,][]{Vasyunin04,Vasyunin08,Wakelam05,Wakelam06,Wakelam10aa,Wakelam10ssr,Penteado17}. In this study, we did not perform full sensitivity analysis of our model to the value of $\epsilon_{\rm ph}$. Instead, we attempted to identify the most likely value of $\epsilon_{\rm ph}$ by comparing the results of numerical simulations with observational data. A full analysis of the sensitivity of the modelling results to the parameters that control grain-surface chemistry is yet to be performed. Variations of some parameters ($E_{\rm b}/E_{\rm D}$, $E_{\rm A,O+CO}$) that significantly affect surface chemistry were briefly investigated in order to identify a physically justified value that yields results, which agree to observations.

In order to acquire an estimate of $\epsilon_{\rm ph}$, we calculated the total differences $\sum (X_{\rm obs}-X_{\rm calc})$ between observed ($X_{\rm obs}$) and calculated ($X_{\rm calc}$) relative abundances along LOS for icy species at corresponding $A_{\rm V}$(LOS) values (i.e., the calculated and observed CO:H$_2$O, CO$_2$:H$_2$O, and NH$_3$:H$_2$O column density ratios). In this manner, we obtained a separate estimate of $\epsilon_{\rm ph}$ from each of the three icy species (carbon oxides and ammonia). This procedure was done only for the valid data points and in the interval of total interstellar extinction $0<A_{\rm V} \rm{(LOS)}\leq 22$ interval for calculations (Sections \ref{31obs} and \ref{32lim}). In Fig.~\ref{att-ices}, the data points used in this analysis can be identified as the black filled circles in the unshaded parts of the plots.

The three `valid' observations of methanol ice were omitted from this analysis  because of the high likelihood that, in most cases, the column densities of interstellar methanol ices towards background stars are too low to be detected, and thus the sample cannot be regarded as representative. According to the modelling results, a major ice component is molecular nitrogen; it could not be included in the analysis because of lack of observations. The number of the `valid' observations was 12 for CO, 9 for CO$_2$ and 11 for NH$_3$.

Table~\ref{tab-epsilo} shows that the total differences are the lowest for $\epsilon_{\rm ph}$ values 0.20, 0.30, and 0.40. Therefore, we propose a value of 0.3 to be used as the ratio between solid and gas phase molecule photodissociation rate coefficients in astrochemical modelling.

\section*{Acknowledgements}

This publication is supported by ERDF project «Physical and chemical processes in the interstellar medium», No 1.1.1.1/16/A/213 being implemented in Ventspils University College. I thank Ventspils City Council for its financial support. I am grateful to Anton Vasyunin for a valuable discussion about this paper. This research has made use of NASA's Astrophysics Data System.

 \footnotesize{
\bibliography{pdi-3f}
\bibliographystyle{mn2e}
 }

\label{lastpage}

\appendix
\section{Observational data employed for comparison with observations.}
\label{app-obs}

Table~\ref{tab-obs} summarizes observations of interstellar ices, considered in this study.

\begin{table*}
 \centering
\caption{Observational data employed in the paper: interstellar extinction along LOS and respective relative abundances $X_{\rm obs}$ of icy species.}
\label{tab-obs}
  \begin{tabular}{@{}lcccccclc@{}}
  \hline
Source ID & Other ID & $A_{\rm V}$(LOS), mag & $\rm \frac{CO}{H_2O}$,\% & $\rm \frac{CO_2}{H_2O}$,\% & $\rm \frac{NH_3}{H_2O}$,\%$^a$ & $\rm \frac{CH_3OH}{H_2O}$,\% & Reference & Notes \\
\hline
 & Elias 3 & 10 & 20.0 & 20 & . . . & . . . & \citet{Bergin05} &  \\
18160600−0225539 &  & 11.0 & . . . & . . . & 17.29 & . . . & \citet{Boogert11} &  \\
21240614+4958310 &  & 14.1 & . . . & . . . & 12.01 & . . . & \citet{Boogert11} &  \\
22063773+5904520 &  & 15.3 & . . . & . . . & 12.70 & . . . & \citet{Boogert11} &  \\
08093135−3604035 &  & 16.6 & . . . & . . . & 16.71 & . . . & \citet{Boogert11} &  \\
18140712−0708413 &  & 16.6 & . . . & . . . & 9.99 & . . . & \citet{Boogert11} &  \\
19214480+1121203 &  & 18.5 & . . . & . . . & 9.65 & . . . & \citet{Boogert11} &  \\
17160860−2058142 &  & 21.6 & . . . & . . . & 4.72 & . . . & \citet{Boogert11} &  \\
18170426−1802408 &  & 22.0 & . . . & . . . & 22.90 & . . . & \citet{Boogert11} &  \\
19201622+1136292 &  & 22.0 & . . . & 38.08 & 3.09 & . . . & \citet{Boogert11} &  \\
17155573−2055312 &  & 23.3 & . . . & . . . & 4.74 & . . . & \citet{Boogert11} &  \\
15421699−5247439 &  & 25.3 & . . . & . . . & 13.45 & . . . & \citet{Boogert11} &  \\
18170429−1802540 &  & 26.0 & . . . & 23.55 & 8.52 & . . . & \citet{Boogert11} &  \\
18165296−1801287 &  & 26.4 & . . . & . . . & 4.48 & . . . & \citet{Boogert11} &  \\
04393886+2611266 &  & 26.4 & . . . & 29.66 & 6.03 & . . . & \citet{Boogert11} &  \\
04215402+1530299 &  & 26.8 & . . . & . . . & 14.80 & . . . & \citet{Boogert11} &  \\
12014598−6508586 &  & 27.0 & . . . & 34.11 & 13.07 & . . . & \citet{Boogert11} &  \\
21240517+4959100 &  & 27.3 & . . . & 37.60 & 6.76 & . . . & \citet{Boogert11} &  \\
19201597+1135146 &  & 27.6 & . . . & . . . & 10.64 & . . . & \citet{Boogert11} &  \\
17160467−2057072 &  & 28.5 & . . . & . . . & 13.99 & . . . & \citet{Boogert11} &  \\
18170957−0814136 &  & 28.8 & . . . & 43.12 & 6.39 & . . . & \citet{Boogert11} &  \\
18165917−1801158 &  & 29.4 & . . . & . . . & 8.76 & . . . & \citet{Boogert11} &  \\
18171366−0813188 &  & 31.5 & . . . & . . . & 9.08 & . . . & \citet{Boogert11} &  \\
17111501−2726180 &  & 34.4 & . . . & . . . & 6.49 & . . . & \citet{Boogert11} &  \\
18170470−0814495 &  & 37.6 & . . . & . . . & 6.13 & . . . & \citet{Boogert11} &  \\
18171181−0814012 &  & 37.9 & . . . & . . . & 7.35 & . . . & \citet{Boogert11} &  \\
08093468−3605266 &  & 39.8 & . . . & . . . & 8.29 & . . . & \citet{Boogert11} &  \\
18172690−0438406 &  & 40.5 & . . . & 43.75 & 7.01 & . . . & \citet{Boogert11} &  \\
12014301−6508422 &  & 45.5 & . . . & . . . & 1.89 & . . . & \citet{Boogert11} &  \\
17112005−2727131 &  & 52.8 & . . . & . . . & 5.71 & . . . & \citet{Boogert11} &  \\
18300061+0115201 &  & 56.1 & . . . & 39.14 & 8.79 & . . . & \citet{Boogert11} &  \\
16022128−4158478 &  & 10.25 & . . . & 58.87 & . . . & . . . & \citet{Boogert13} & $^b$ \\
16021102−4158468 &  & 11.05 & . . . & 187.8 & . . . & . . . & \citet{Boogert13} & $^b$ \\
16004925−4150320 &  & 14.85 & . . . & 28.08 & . . . & . . . & \citet{Boogert13} & $^b$ \\
16012825−4153521 &  & 24.25 & . . . & 80.32 & . . . & . . . & \citet{Boogert13} & $^b$ \\
15452747−3425184 &  & 24.3 & 25.44 & 42.74 & 2.5 & . . . & \citet{Boogert13} & $^b$ \\
16012635−4150422 &  & 25.35 & . . . & . . . & 10.58 & . . . & \citet{Boogert13} & $^b$ \\
15423699−3407362 &  & 27.6 & . . . & . . . & 6.01 & . . . & \citet{Boogert13} & $^b$ \\
16010642−4202023 &  & 29.45 & . . . & 99.8 & . . . & . . . & \citet{Boogert13} & $^b$ \\
16004739−4203573 &  & 31.25 & . . . & 71.88 & . . . & . . . & \citet{Boogert13} & $^b$ \\
16014254−4153064 &  & 37.8 & 43.1 & 32.14 & 7.14 & . . . & \citet{Boogert13} & $^b$ \\
21443293+4734569 & Q22-1 & 17.49 & . . . & . . . & 3.5 & 1.1 & \citet{Chiar11} &  \\
21461164+4734542 & Q21-6 & 20.7 & . . . & . . . & 3 & 1.6 & \citet{Chiar11} &  \\
 & Elias 25 & 11 & 6 & . . . & . . . & . . . & \citet{Shuping00} & $^c$ \\
 & Elias 32 & 24 & 22 & . . . & . . . & . . . & \citet{Shuping00} & $^c$ \\
 & Elias 33 & 31 & 36 & . . . & . . . & . . . & \citet{Shuping00} & $^c$ \\
 & WL 16 & 32 & 4 & . . . & . . . & . . . & \citet{Shuping00} & $^c$ \\
 & WL 12 & 41 & 10 & . . . & . . . & . . . & \citet{Shuping00} & $^c$ \\
 & Elias 29 & 48 & 5 & . . . & . . . & . . . & \citet{Shuping00} & $^c$ \\
 & WL 6 & 48 & 21 & . . . & . . . & . . . & \citet{Shuping00} & $^c$ \\
 & WL 5 & 50 & 15 & . . . & . . . & . . . & \citet{Shuping00} & $^c$ \\
042324.6+250009 & Elias 3 & 8.4 & 28.7 & . . . & . . . & . . . & \citet{Teixeira99} &  \\
043325.9+261534 & Elias 13 & 11.6 & 11.8 & . . . & . . . & . . . & \citet{Teixeira99} &  \\
043926.8+255259 & Elias 15 & 15 & 27.1 & . . . & . . . & . . . & \citet{Teixeira99} &  \\
044058.1+255416 & Tamura 8 & 21.5 & 26.0 & . . . & . . . & . . . & \citet{Teixeira99} &  \\
043938.7+261125 & Elias 16 & 23.5 & 33.6 & . . . & . . . & . . . & \citet{Teixeira99} &  \\
043728.2+261024 & Tamura 2 & 6.3 & . . . & 25.0 & . . . & . . . & \citet{Whittet07} &  \\
043325.9+261534 & Elias 13 & 11.7 & 12.0 & 19.4 & . . . & . . . & \citet{Whittet07} &  \\
043926.9+255259 & Elias 15 & 15.3 & 27.3 & 16.7 & . . . & . . . & \citet{Whittet07} &  \\
042630.7+243637 &  & 17.8 & 45.1 & 18.3 & . . . & . . . & \citet{Whittet07} &  \\
043213.2+242910 &  & 20.9 & 33.3 & 16.0 & . . . & . . . & \citet{Whittet07} &  \\
044057.5+255413 & Tamura 8 & 21.5 & 24.3 & . . . & . . . & . . . & \citet{Whittet07} &  \\
043938.9+261125 & Elias 16 & 24.1 & 25.3 & 21.0 & . . . & . . . & \citet{Whittet07} &  \\
\end{tabular}\\
\end{table*}

\begin{table*}
 \centering
\caption{Table~\ref{tab-obs} continued.}
\label{tab-obs1}
  \begin{tabular}{@{}lcccccclc@{}}
  \hline
Source ID & Other ID & $A_{\rm V}$(LOS), mag & $\rm \frac{CO}{H_2O}$,\% & $\rm \frac{CO_2}{H_2O}$,\% & $\rm \frac{NH_3}{H_2O}$,\%$^a$ & $\rm \frac{CH_3OH}{H_2O}$,\% & Reference & Notes \\
\hline
J04332594+2615334 & Elias 13 & 11.7 & 8.3 & 16.7 & . . . & . . . & \citet{Whittet11} &  \\
J18140712−0708413 &  & 15.1 & . . . & . . . & . . . & 5.3 & \citet{Whittet11} & $^d$ \\
J19201622+1136292 &  & 20.0 & . . . & 38.2 & . . . & . . . & \citet{Whittet11} &  \\
J18170429−1802540 &  & 23.7 & . . . & 23.8 & . . . & . . . & \citet{Whittet11} &  \\
J21472204+4734410 & Q21-1 & 23.9 & 27.0 & 34.5 & . . . & 0.9 & \citet{Whittet11} &  \\
J12014598−6508586 &  & 24.6 & . . . & 34.0 & . . . & . . . & \citet{Whittet11} &  \\
J21240517+4959100 &  & 24.8 & . . . & 37.6 & . . . & 7.0 & \citet{Whittet11} & $^d$ \\
J18170957−0814136 &  & 26.2 & . . . & 43.2 & . . . & 10.0 & \citet{Whittet11} & $^d$ \\
J18171366−0813188 &  & 28.6 & . . . & . . . & . . . & 10.9 & \citet{Whittet11} & $^d$ \\
J18300061+0115201 & CK2, EC 118 & 34 & 46.3 & 28.3 & . . . & . . . & \citet{Whittet11} &  \\
J18170470−0814495 &  & 34.2 & . . . & . . . & . . . & 6.8 & \citet{Whittet11} & $^d$ \\
J18171181−0814012 &  & 34.5 & . . . & . . . & . . . & 12.3 & \citet{Whittet11} & $^d$ \\
J18172690−0438406 &  & 36.8 & . . . & 43.9 & . . . & 8.9 & \citet{Whittet11} & $^d$ \\
J15542044−0254073 &  & 12.5 & 40.4 &  & . . . & . . . & \citet{Whittet13} &  \\
\hline
\end{tabular}\\
$^a$\,Includes $\rm NH_4^+$.
$^b$\,Lupus cloud; increased irradiation.
$^c$\,$\rho$ Oph cloud; strong irradiation.
$^d$\,Identified as long-lived stable starless cloud core.
\end{table*}

\section{Whole-grain heating frequency by cosmic rays.}
\label{app-crd}
%
\begin{figure}
 \vspace{-2.0cm}
 \hspace{-1.5cm}
  \includegraphics[width=19.0cm]{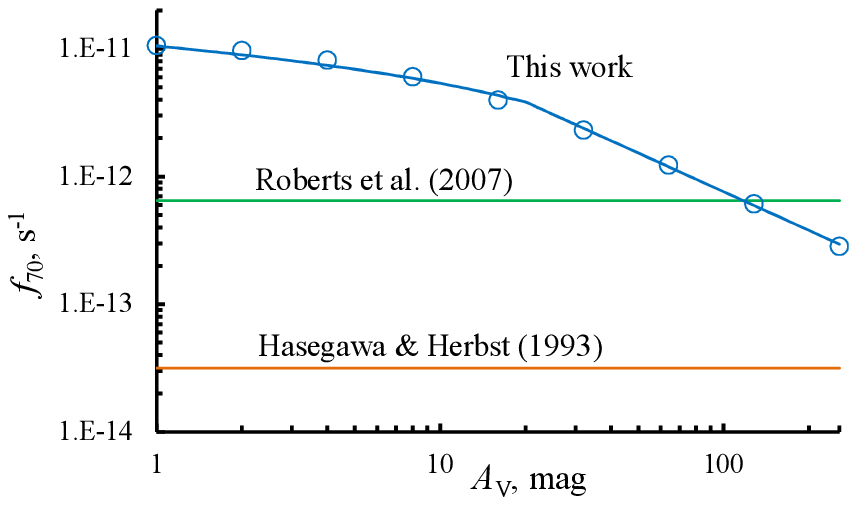}
 \vspace{-21.0cm}
 \caption{Calculated frequency of CR encounters with interstellar grains resulting in grain temperatures in excess of 70~K. For comparison, the figure shows frequencies derived by previous studies.}
 \label{att-f70}
\end{figure}
Our task is to obtain the frequency of CR encounters with interstellar grains that heat the grains to 70~K ($f_{70}$, s$^{-1}$) as a function of column density. For convenience, the latter is expressed in terms of interstellar extinction $A_{\rm V}=N_{\rm H}/(2.0\times10^{21})$, where $N_{\rm H}$ (cm$^{-2}$) is the total hydrogen column density from cloud edge to the modelled spatial point. From the grain types considered by \citet{K16}, we take 0.1\,$\mu$m grains with a thin 0.01\,$\mu$m ice layer at an equilibrium temperature of 10\,K. This is the type of grains most relevant to this study, where the accumulation of the icy mantles in a quiescent medium is considered. \citet{K16} provides grain heating frequencies also for bare grains an grains with a 0.03\,$\mu$m thick ice layer.

The following procedure was used to acquire a reasonable estimate of $f_{70}$ as a function of $A_{\rm V}$. First, from the data of \citet{K16} we identify that iron ions with an energy of 26~MeV\,amu$^{-1}$ are the CR particles that contribute most to grain heating to a temperature of 70~K. Second, with the methodology described by \citet{Chabot16} and \citet{K16}, we calculated the flux of these ions at $A_{\rm V}$ values equal to $2^x$\,mag, where $x$ is -1; 0; 1; 2; 3; 4; 5; 6; 7 and 8 (i.e., 0.5..256\,mag). Third, we assumed that all the CR ions contributing to grain heating to 70~K are absorbed in the cloud in a manner similar to the 26~MeV\,amu$^{-1}$ Fe ions. In other words, the 26~MeV\,amu$^{-1}$ Fe ions dominate grain heating to 70~K. Thus, a set of source data points were obtained.

These points imply a rather high rate for whole-grain heating, corresponding to a CR spectrum with abundant low-energy ($<100$~MeV\,amu$^{-1}$) particles \citep{Moskalenko02}. However, two aspects limit the grain heating rate. First, it has been shown that this type of CR spectra has its intensity effectively reduced by relatively shallow hydrogen column densities \citep{Padovani09}. For example, the intensity of all CR species is reduced by almost two orders of magnitude between H$_2$ column densities of $10^{21}$~cm$^{-2}$ and $10^{23}$~cm$^{-2}$ \citep{Chabot16}. Because of this, it is necessary to adopt the same geometrical setting we employed for the ISRF, i.e., CRs are allowed to enter the cloud core only from two opposite sides of the cloud. This reduces the values of $f_{70}$ data points by a factor of $\approx1/(2\pi)$ \citep[][considered isotropic irradiation from all sides, i.e., spherical geometry]{K16}. Second, similarly to the case of $\zeta$ (Section~\ref{21phys}), we adopt a factor of 0.3 that modifies the CR flux because of magnetic reflection.

$A_{\rm V}$-dependent analytical function of $f_{70}$ was then obtained by interpolating the acquired data points. The grains are heated by CRs to a temperature of 70~K with the frequency $f_{70}$ for the interval $0.5\leq A_{\rm V}<20$:
	\begin{equation}
   \label{crd1}
f_{70} = -2.246\times 10^{-12}\rmn{ln}(A_{\rm V})+1.055\times 10^{-11}; \\
  \end{equation}
and for $20\leq A_{\rm V}\leq256$~mag:
	\begin{equation}
   \label{crd2}
f_{70} = 7.826\times 10^{-11}A_{\rm V}^{-1.006}.
  \end{equation}
The resulting function is shown graphically in Figure~\ref{att-f70}. To calculate the desorption rate for species $i$, $f_{70}$ must be used according to Equation\,(\ref{chem1}) in the paper.

%
\section{Model for photodissociative desorption.}
\label{app-pdd}
%
\begin{figure}
 \vspace{-3.0cm}
  \hspace{-2.0cm}
  \includegraphics[width=18.0cm]{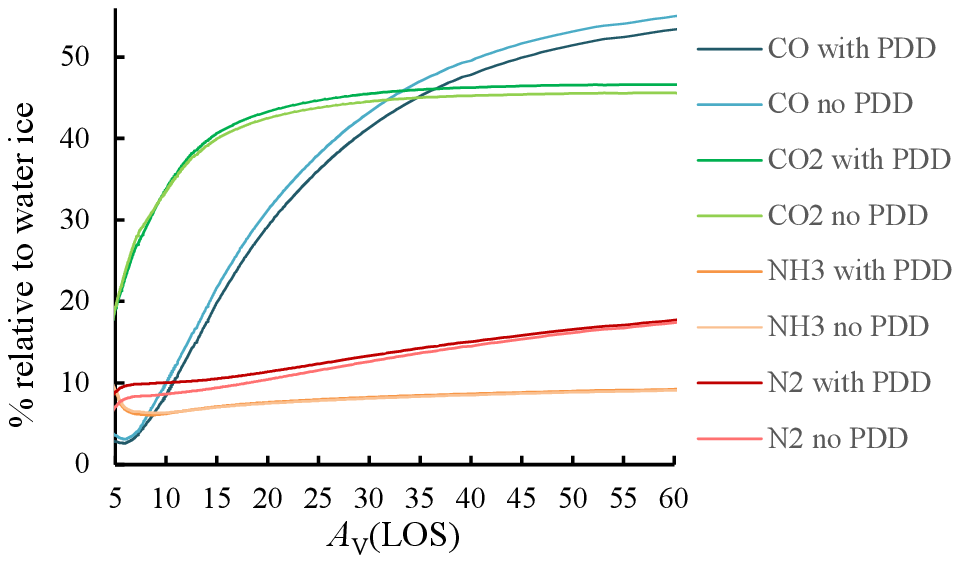}
	\vspace{-18.0cm}
 \caption{Comparison of calculated relative abundances of abundant icy species for a model with and without photodissociative desorption (see text). For maximum effect, photodissociation efficiency $\epsilon_{\rm ph}$ was taken to be 1.0.}
 \label{att-pdd}
\end{figure}
Dissociation of surface water results in gas-phase products containing the oxygen atom (i.e., PDD) with a rate that is equal to 3 per cent of the total dissociation rate \citep{Andersson08}. From this we assume that 3 percent of all photodissociated surface species are desorbed in the model. The photodissociation coefficient (total) was calculated with Equation\,(\ref{chem2}) with $B=0$. As a rule, the PDD rate never exceeds the total \textit{photodesorption} rate. This means that the photodesorption rate (coefficient calculated with Equation~(\ref{chem3})) accounts for both, the desorption of intact and dissociated species. This approach on PDD is rather conservative; however, it is an improvement over our previous models that did not consider PDD at all.

The adopted approach results in that CO and N$_2$ are desorbed intact for most of the time, while other important species are largely removed via PDD, which is in general agreement with experiments \citep[e.g.,][]{Oberg09apj,Oberg09aa1,Fayolle11,Fayolle13,Bertin12,Bertin16,Fillion14,Cruz16}. Because the dissociation rate coefficients tend to decrease more rapidly with $A_{\rm V}$, the effects of PDD diminish with cloud evolution in our model. For example, at $A_{\rm V}(LOS)=5$~mag, 99.8 per cent of CO and 50.4 per cent of H$_2$O molecules are desorbed intact by ISRF photons. At $A_{\rm V}(LOS)=50$~mag, these numbers are 100.0 and 98.9, respectively. Therefore, the PDD rates do not become overestimated. Figure~\ref{att-pdd} shows that the effects of PDD on the relative abundances of major icy species are relatively negligible for this study. They are even less pronounced for simulations with lower values of $\epsilon_{\rm ph}$ (i.e., lower icy molecule photodissociation rates).

The rate for water photodissociation by the ISRF at $A_{\rm V}$ values below 2~mag can be close to $10^{-16}$\,s$^{-1}$cm$^{-3}$. This is comparable to the rate of free H atoms sticking on grain surfaces. Therefore, the photolysis of hydrogen-containing molecules might serve as an important source of atomic hydrogen on the surface. However, this is likely not the case. Most of H atoms, split-off from surface molecules, leave the grain \citep{Andersson06}. To reflect this aspect in the model, we assumed that 95 per cent of all H atoms (and H$_2$ molecules) split off from surface H$_2$O are desorbed \citep{Arasa15}. This rule was attributed also to the two other major hydrogenated surface species ammonia NH$_3$ and methanol CH$_3$OH.

\end{document}